\documentclass{emulateapj}

\def\FLASH                 {{\sc flash}}
\def\PARAMESH              {{\sc paramesh}}
\def\mag{{\bf B}}
\def\mach{{\cal M}}
\def\gsim{\;\lower4pt\hbox{${\buildrel\displaystyle >\over\sim}$}\,}
\def\lsim{\;\lower4pt\hbox{${\buildrel\displaystyle <\over\sim}$}\,}

\shorttitle{Anisotropic thermal conduction in SNRs}
\shortauthors{Orlando et al.}

\begin{document}


\title{The importance of magnetic-field-oriented thermal conduction
          in the interaction of SNR shocks with interstellar clouds}

\author{S. Orlando\altaffilmark{1},
        F. Bocchino\altaffilmark{1}}
\affil{INAF - Osservatorio Astronomico di Palermo ``G.S.
       Vaiana'', Piazza del Parlamento 1, 90134 Palermo, Italy}
\author{F. Reale\altaffilmark{2,1},
        G. Peres\altaffilmark{2,1} and
        P. Pagano}
\affil{Dip. di Scienze Fisiche \& Astronomiche, Univ. di
       Palermo, Piazza del Parlamento 1, 90134 Palermo,
       Italy}


\altaffiltext{1}{Consorzio COMETA, via Santa Sofia 64, 95123 Catania, Italy}
\altaffiltext{2}{INAF, Viale del Parco Mellini 84, 00136 Roma, Italy}


\begin{abstract}
We explore the importance of magnetic-field-oriented thermal conduction
in the interaction of supernova remnant (SNR) shocks with radiative gas
clouds and in determining the mass and energy exchange between the clouds
and the hot surrounding medium. We perform 2.5D MHD simulations of a shock
impacting on an isolated gas cloud, including anisotropic thermal
conduction and radiative cooling; we consider the representative case of
a Mach 50 shock impacting on a cloud ten-fold denser than the ambient
medium. We consider different configurations of the ambient magnetic
field and compare MHD models with or without the thermal conduction. The
efficiency of the thermal conduction in the presence of magnetic field is,
in general, reduced with respect to the unmagnetized case. The reduction
factor strongly depends on the initial magnetic field orientation,
and it is minimum when the magnetic field is initially aligned with the
direction of shock propagation. The thermal conduction contributes to
suppress hydrodynamic instabilities, reducing the mass mixing of the
cloud and preserving the cloud from complete fragmentation. Depending
on the magnetic field orientation, the heat conduction may determine a
significant energy exchange between the cloud and the hot surrounding
medium which, while remaining always at levels less than those in the
unmagnetized case, leads to a progressive heating and evaporation of
the cloud. This additional heating may contrast the radiative cooling
of some parts of the cloud, preventing the onset of thermal instabilities.
\end{abstract}




\section{Introduction}
\label{sec1}

The interaction of the shock waves of supernova remnants (SNRs) with the
magnetized and inhomogeneous interstellar medium (ISM) is responsible of
the great morphological complexity of SNRs and certainly plays a major
role in determining the exchange of mass, momentum, and energy between
diffuse hot plasma and dense clouds or clumps. These exchanges may occur
through, for example, hydrodynamic ablation and thermal conduction and,
among other things, lead to the cloud crushing and to the reduction of
the Jeans mass causing star formation.

The propagation of hot SNR shock fronts in the ISM and their interaction
with local over-dense gas clouds have been investigated with detailed
hydrodynamic and MHD modeling. The most complete review of this problem
in the unmagnetized, non-conducting, and non-radiative limits is provided
by \cite{1994ApJ...420..213K}. These studies have shown that the cloud is
disrupted by the action of both Kelvin-Helmholtz (KH) and Rayleigh-Taylor
(RT) instabilities after several crushing times, with the cloud material
expanding and diffusing into the ambient medium. An ambient magnetic
field can both act as a confinement mechanism of the plasma and be
modified by the interstellar flow and by local field stretching. Also,
a strong magnetic field is known to limit hydrodynamic instabilities
developing during the shock-cloud interaction by providing an additional
tension at the interface between the cloud and the surrounding medium
\citep[e.g][]{1994ApJ...433..757M, 1996ApJ...473..365J}.

The interaction of the shock with a {\it radiative} cloud has been
only recently analyzed in detail \citep[e.g.][]{2002A&A...395L..13M,
2004ApJ...604...74F}. 2D calculations have shown that the effect of
the radiative cooling is to break up the clouds into numerous dense
and cold fragments that survive for many dynamical timescales. In the
case of the interaction between magnetized shocks and radiative clouds,
the magnetic field may enhance the efficiency of the radiative cooling,
influencing the size and distribution of condensed cooled fragments
\citep{2005ApJ...619..327F}.

The role played by the thermal conduction during the shock-cloud
interaction has been less studied so far. In a previous paper,
\cite{2005A&A...444..505O} (hereafter Paper I) have addressed this point
in the unmagnetized limit. In particular, we have investigated the effect
of thermal conduction and radiative cooling on the cloud evolution and on
the mass and energy exchange between the cloud and the surrounding medium;
we have selected and explored two different physical regimes chosen so
that either one of the processes is dominant. In the case dominated by
the radiative losses, we have found that the shocked cloud fragments into
cold, dense, and compact filaments surrounded by a hot corona which is
ablated by the thermal conduction. Instead, in the case dominated by
thermal conduction, the shocked cloud evaporates in a few dynamical
timescales. In both cases, we have found that the thermal conduction
is very effective in suppressing the hydrodynamic instabilities that
would develop at the cloud boundaries, preserving the cloud from complete
destruction. \cite{orlando2} and \cite{2006A&A...458..213M} have studied
the observable effects of thermal conduction on the evolution of the
shocked cloud in the X-ray band.

Here, we extend the previous studies by investigating the effect of
the thermal conduction in a magnetized medium, unexplored so far.
Of special interest to us is to investigate the role of anisotropic
thermal conduction - funneled by locally organized magnetic fields -
in the mass and energy exchange between ISM phases. In particular,
we aim at addressing the following questions: How and under which
physical conditions does the magnetic-field-oriented thermal conduction
influence the evolution of the shocked cloud? How do the mass mixing
of the cloud material and the energy exchange between the cloud and the
surrounding medium depend on the orientation and strength of the magnetic
field and on the efficiency of the thermal conduction? 

To answer these questions, we take as representative the model case of
a shock with Mach number $\mach = 50$ (corresponding to a post-shock
temperature $T\approx 4.7\times 10^6$ K for an unperturbed medium
with $T=10^4$ K) impacting on an isolated cloud ten-fold denser than
the ambient medium. Paper I has shown that, in this case, the thermal
conduction dominates the evolution of the shocked cloud in the absence of
magnetic field. Around this basic configuration, we perform a set of MHD
simulations, with different interstellar magnetic field orientations,
and compare models calculated with thermal conduction turned either
``on'' or ``off'' in order to identify its effects on the cloud evolution.

The paper is organized as follows: in Sect. \ref{sec2} we describe the
MHD model and the numerical setup; in Sect. \ref{sec3} we discuss the
results; and finally in Sect. \ref{sec4} we draw our conclusions.

%
\section{The model}
\label{sec2}

We model the impact of a planar supernova shock front onto an isolated
gas cloud. The shock propagates through a magnetized ambient medium and
the cloud is assumed to be small compared to the curvature radius of
the shock\footnote{In the case of a small cloud, the SNR does not evolve
significantly during the shock-cloud interaction, and the assumption of a
planar shock is justified \citep[see also][]{1994ApJ...420..213K}.}.
The fluid is assumed to be fully ionized with a ratio of specific
heats $\gamma = 5/3$. The model includes radiative cooling, thermal
conduction (including the effects of heat flux saturation) and resistivity
effects. The shock-cloud interaction is modeled by solving numerically
the time-dependent non-ideal MHD equations (written in non-dimensional
conservative form):

\begin{equation}
\frac{\partial \rho}{\partial t} + \nabla \cdot (\rho {\bf u}) = 0~,
\end{equation}

\begin{equation}
\frac{\partial \rho {\bf u}}{\partial t} + \nabla \cdot (\rho
{\bf uu}-{\bf BB}) + \nabla P_* = 0~,
\end{equation}

\begin{eqnarray}
\lefteqn{\frac{\partial \rho E}{\partial t} +\nabla\cdot [{\bf u}(\rho
E+P_*) -{\bf B}({\bf u}\cdot {\bf B})] =} \nonumber \\
 & \displaystyle ~~~~~~~~~~~~~~~
\nabla\cdot [{\bf B}\times(\eta\nabla\times{\bf B})]
-\nabla\cdot {\bf F}_{\rm c} -n_{\rm e} n_{\rm H} \Lambda(T)~,
\end{eqnarray}

\begin{equation}
\frac{\partial {\bf B}}{\partial t} +\nabla \cdot({\bf uB}-{\bf
Bu}) = -\nabla\times(\eta\nabla\times{\bf B})~,
\end{equation}

\noindent
where

\[
P_* = P + \frac{B^2}{2}~,~~~~~~~~~~~~~
E = \epsilon +\frac{1}{2} u^2+\frac{1}{2}\frac{B^2}{\rho}~,
\]

\noindent
are the total pressure, and the total gas energy (internal energy,
$\epsilon$, kinetic energy, and magnetic energy) respectively, $t$
is the time, $\rho = \mu m_H n_{\rm H}$ is the mass density, $\mu =
1.26$ is the mean atomic mass (assuming cosmic abundances), $m_H$
is the mass of the hydrogen atom, $n_{\rm H}$ is the hydrogen number
density, {\bf u} is the gas velocity, $T$ is the temperature, {\mag} is
the magnetic field, $\eta$ is the resistivity according to \cite{spi62},
${\bf F}_{\rm c}$ is the conductive flux, and $\Lambda(T)$ represents the
radiative losses per unit emission measure \citep[e.g.][]{rs77, mgv85,
2000adnx.conf..161K}. We use the ideal gas law, $P=(\gamma-1) \rho
\epsilon$.

In order to track the original cloud material, we use a tracer that
is passively advected in the same manner as the density. We define
$C_{\rm cl}$ the mass fraction of the cloud inside the computational
cell. The cloud material is initialized with $C_{\rm cl} = 1$, while
$C_{\rm cl} = 0$ in the ambient medium\footnote{We checked that the
used numerical scheme ensures that always $0 \leq C_{\rm cl} \leq 1$.}.
During the shock-cloud evolution, the cloud and the ambient medium mix
together, leading to regions with $0 < C_{\rm cl} < 1$. At any time $t$
the density of cloud material in a fluid cell is given by $\rho_{\rm cl}
= \rho C_{\rm cl}$.

The thermal conductivity in an organized magnetic field is known to be
highly anisotropic and it can be extraordinarily reduced in the direction
transverse to the field. The thermal flux, therefore, is locally split
into two components, along and across the magnetic field lines, 
${\bf F}_{\rm c} = F_{\parallel}~{\bf i}+F_{\perp}~{\bf j}$, where

\begin{equation}
\begin{array}{l}\displaystyle
F_{\parallel} = \left(\frac{1}{[q_{\rm spi}]_{\parallel}}+
                \frac{1}{[q_{\rm sat}]_{\parallel}}\right)^{-1}~,
\\ \\ \displaystyle
F_{\perp} = \left(\frac{1}{[q_{\rm spi}]_{\perp}}+
               \frac{1}{[q_{\rm sat}]_{\perp}}\right)^{-1}~,
\end{array}
\label{cond}
\end{equation}

\noindent
to allow for a smooth transition between the classical and saturated
conduction regime. In Eqs. \ref{cond}, $[q_{\rm spi}]_{\parallel}$ and
$[q_{\rm spi}]_{\perp}$ represent the classical conductive flux along
and across the magnetic field lines \citep{spi62}

\begin{equation}
\begin{array}{l}\displaystyle
[q_{\rm spi}]_{\parallel} = -\kappa_{\parallel} [\nabla T]_{\parallel}
\approx - 5.6\times 10^{-7} T^{5/2}~ [\nabla T]_{\parallel}
\\ \\ \displaystyle
[q_{\rm spi}]_{\perp} = -\kappa_{\perp} [\nabla T]_{\perp}
\approx - 3.3\times 10^{-16} \frac{n^2_{\rm H}}{T^{1/2}B^2}~ [\nabla
T]_{\perp}
\end{array}
\label{spit_eq}
\end{equation}

\noindent
where $[\nabla T]_{\parallel}$ and $[\nabla T]_{\perp}$ are the thermal
gradients along and across the magnetic field, and $\kappa_{\parallel}$ and
$\kappa_{\perp}$ (in units of erg s$^{-1}$ K$^{-1}$ cm$^{-1}$) are the
thermal conduction coefficients along and across the magnetic field
lines\footnote{For the values of $T$, $n_{\rm H}$ and $B$ used here,
$\kappa_{\parallel}/\kappa_{\perp}\approx 10^{16}$ at the beginning of
the shock-cloud interaction.}, respectively. The saturated flux along
and across the magnetic field lines, $[q_{\rm sat}]_{\parallel}$ and
$[q_{\rm sat}]_{\perp}$, are \citep{cm77}

\begin{equation}
\begin{array}{l}\displaystyle
[q_{\rm sat}]_{\parallel} = -\mbox{sign}\left([\nabla T]_{\parallel}\right)~ 
                5\phi \rho c_{\rm s}^3,
\\ \\ \displaystyle
[q_{\rm sat}]_{\perp} = -\mbox{sign}\left([\nabla T]_{\perp}\right)~ 
                5\phi \rho c_{\rm s}^3,
\end{array}
\label{therm}
\end{equation}

\noindent
where $c_{\rm s}$ is the isothermal sound speed, and $\phi$ is a number
of the order of unity; we set $\phi = 0.3$ according to the values
suggested for a fully ionized cosmic gas: $0.24<\phi<0.35$ \citep[][
and references therein]{1984ApJ...277..605G, 1989ApJ...336..979B,
2002A&A...392..735F}. As discussed in Paper I, this choice implies that
no thermal precursor develops during the shock propagation, consistent
with the fact that no precursor is observed in young and middle aged SNRs.

The initial unperturbed ambient medium is magnetized, isothermal
(with temperature $T_{\rm ism}=10^4$ K, corresponding to an isothermal
sound speed $c_{\rm ism} = 11.5$ km s$^{-1}$), and uniform (with hydrogen
number density $n_{\rm ism} = 0.1$ cm$^{-3}$; see Table \ref{tab1}).
The gas cloud is in pressure equilibrium with its surrounding and has
a circular cross-section with radius $r_{\rm cl} = 1$~pc; its radial
density distribution is given by

\begin{equation}
n_{\rm cl}(r) = n_{\rm ism}+\frac{n_{\rm cl0}-n_{\rm ism}}
{\cosh\left[\sigma\left(r/r_{\rm cl}\right)^{\sigma}\right]}~~~,
\end{equation}

\noindent
where $n_{\rm cl0}$ is the hydrogen number density at the cloud center, $r$
is the radial distance from the cloud center and $\sigma=10$. The above
distribution describes a thin transition layer ($\sim 0.3~r_{\rm cl}$)
around the cloud that smoothly brings the cloud density to the value
of the surrounding medium\footnote{A finite transition layer, in
general, is expected in real interstellar clouds due, for instance,
to thermal conduction (\citealt{1986ApJ...304..787B}; see also
\citealt{2006ApJS..164..477N}).}. The initial density contrast between
the cloud center and the ambient medium is $\chi=n_{\rm cl0}/n_{\rm ism}
= 10$. The cloud temperature is determined by the pressure balance
across the cloud boundary.

\begin{deluxetable}{llll}
\footnotesize
\tablecaption{Summary of the initial physical parameters characterizing the
MHD simulations. \label{tab1}}
\tablewidth{0pt}
\tablehead{
\colhead{ } & \colhead{Temperature}   & \colhead{Density}   & \colhead{Velocity}
}
\startdata
{\it ISM} & $10^4$ K & $0.1$ cm$^{-3}$ & 0.0 \\
{\it Cloud} & $10^3$ K & $1.0$ cm$^{-3}$ & 0.0 \\
{\it Post-shock medium:} & $4.7\times 10^6$ K & $0.4$ cm$^{-3}$ & $430$
km s$^{-1}$ \\
\enddata
\end{deluxetable}

The SNR shock front propagates with a velocity $w= \mach c_{\rm ism}$
in the ambient medium, where $\mach$ is the shock Mach number, and
$c_{\rm ism}$ is the sound speed in the interstellar medium; we consider
a shock propagating with $\mach = 50$, i.e. a shock velocity $w\approx
570$ km s$^{-1}$ and a temperature $T_{psh}\approx 4.7\times 10^6$~K.
As discussed in Paper I, in this case (for a cloud with $r_{\rm cl} =
1$~pc and $\chi=10$) the cloud dynamics would be dominated by thermal
conduction in the absence of magnetic field. The post-shock conditions
of the ambient medium well before the impact onto the cloud are given
by the strong shock limit \citep{zel66}.

Starting from this basic configuration, we consider a set of simulations
with different initial magnetic field orientations. We adopt a 2.5D
Cartesian coordinate system $(x,y)$, implying that the simulated clouds
are cylinders extending infinitely along the $z$ axis perpendicular to
the $(x,y)$ plane. The primary shock propagates along the $y$ axis.
In this geometry, we consider three different field orientations: 1)
parallel to the planar shock and perpendicular to the cylindrical cloud,
2) perpendicular to both the shock front and the cloud, and 3) parallel
to both the shock and the cloud. The magnetic field components along the
$x$ and the $z$ axis are enhanced by a factor $(\gamma+1)/(\gamma-1)$
(where $\gamma$ is the ratio of specific heats) in the post-shock region
\citep[in the strong shock limit;][]{zel66}, whereas the component
along the $y$ axis is continuous across the shock. We include runs in
the strong and weak magnetic field limits, considering initial field
strengths of $|\mag| = 2.63,\,1.31,\,0.26,\,0~\mu$G in the unperturbed
ambient medium\footnote{The unmagnetized case (i.e. $|\mag| =0$)
described here is analogous to the one studied in Paper I except for
the fact that in the present case the cloud is a cylinder rather than
a sphere and has smooth boundaries.}, corresponding to $\beta_0 =
1,\,4,\,100,\,\infty$, where $\beta_{\rm 0} = P/(B^2/8\pi)$ is the
ratio of thermal to magnetic pressure in the pre-shock region. This
range of $\beta_{\rm 0}$ includes typical values inferred for the
diffuse regions of the ISM \citep[e.g.][]{2004RvMP...76..125M}
and for shock-cloud interaction regions in evolved SNR shells
\citep[e.g][]{2000A&A...359..316B}. There is no magnetic field component
exclusively associated to the cloud.

We follow the shock-cloud interaction for $3.5\,\tau_{\rm cc}$, where
$\tau_{\rm cc}\approx \chi^{1/2} r_{\rm cl}/w$ is the cloud crushing time,
i.e. the characteristic time of the shock transmission through the
cloud; for the conditions considered here ($\chi=10$ and $\mach = 50$),
$\tau_{\rm cc} \approx 5.4\times 10^3$ yr. Each simulation is repeated
either with or without thermal conduction for each field orientation.
Table \ref{tab2} lists the runs and the initial physical parameters of
the simulations.

\begin{deluxetable}{lcccccc}
\footnotesize
\tablecaption{Summary of the MHD simulations. In all runs the shock Mach
    number is $\mach=50$, the density contrast is $\chi=10$, and the
    cloud crushing time is $\tau_{\rm cc} \approx 5.4\times 10^3$ yr.
    \label{tab2}}
\tablewidth{0pt}
\tablehead{
\colhead{Run} & \colhead{$|\mag|$} & \colhead{$\beta_0$} & \colhead{Field} &
\colhead{Therm.} & \colhead{Rad.} & \colhead{Res.$^a$}\\
\colhead{ } & \colhead{$\mu$G} & \colhead{ } & \colhead{Comp.} &
\colhead{Cond.} & \colhead{Losses} \\
}
\startdata
   NN       & 0    & $\infty$  & $-$ & no & no & 132 \\
   NR       & 0    & $\infty$  & $-$ & no & yes & 132 \\
   TN       & 0    & $\infty$  & $-$ & yes & no & 132 \\
   TR       & 0    & $\infty$  & $-$ & yes & yes & 132 \\
   NN-Bx4   & 1.31 & 4   & $B_{\rm x}$ & no & no & 132 \\
   NN-By4   & 1.31 & 4   & $B_{\rm y}$ & no & no & 132 \\
   NN-Bz4   & 1.31 & 4   & $B_{\rm z}$ & no & no & 132 \\
   TN-Bx4   & 1.31 & 4   & $B_{\rm x}$ & yes & no & 132 \\
   TN-By4   & 1.31 & 4   & $B_{\rm y}$ & yes & no & 132 \\
   TN-Bz4   & 1.31 & 4   & $B_{\rm z}$ & yes & no & 132 \\
   TR-Bx1   & 2.63 & 1   & $B_{\rm x}$ & yes & yes & 132 \\
   TR-By1   & 2.63 & 1   & $B_{\rm y}$ & yes & yes & 132 \\
   TR-Bz1   & 2.63 & 1   & $B_{\rm z}$ & yes & yes & 132 \\
   TR-Bx4   & 1.31 & 4   & $B_{\rm x}$ & yes & yes & 132 \\
   TR-By4   & 1.31 & 4   & $B_{\rm y}$ & yes & yes & 132 \\
   TR-Bz4   & 1.31 & 4   & $B_{\rm z}$ & yes & yes & 132 \\
   TR-Bx100 & 0.26 & 100 & $B_{\rm x}$ & yes & yes & 132 \\
   TR-By100 & 0.26 & 100 & $B_{\rm y}$ & yes & yes & 132 \\
   TR-Bz100 & 0.26 & 100 & $B_{\rm z}$ & yes & yes & 132 \\
\\
   TR-Bz4-hr & 1.31 & 4   & $B_{\rm z}$ & yes & yes & 264 \\
   TR-Bz4-hr2 & 1.31 & 4   & $B_{\rm z}$ & yes & yes & 528 \\
\enddata
\tablenotetext{a}{Initial number of zones per cloud radius}
\end{deluxetable}

\begin{figure*}
  \centering
  \epsscale{1.07}
  \plotone{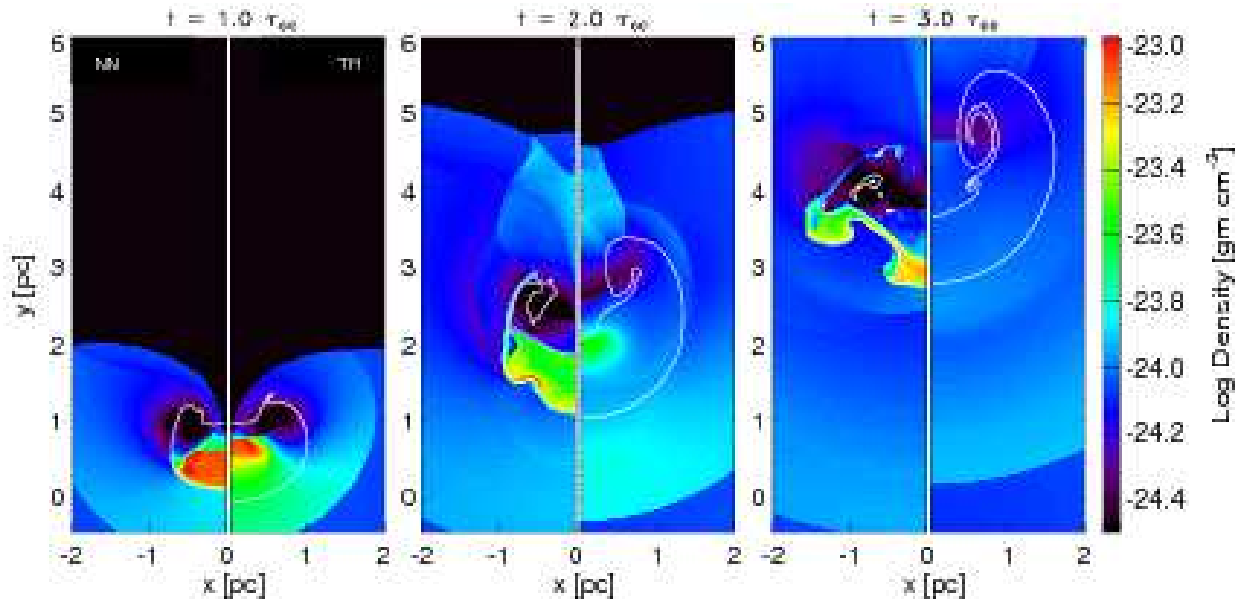}
  \caption{Mass density distribution (gm cm$^{-3}$) in the $(x,y)$
   plane, in log scale, in the simulations NN (left half panels) and
   TR (right half panels), sampled at the labeled times in units of
   $\tau_{\rm cc}$. The contour encloses the cloud material.}
   \label{fig1}
\end{figure*}

We solve numerically the set of MHD equations using \FLASH\
\citep{for00}, a multiphysics code including the \PARAMESH\
library \citep{mom00} for the adaptive mesh refinement. The MHD
equations are solved using the \FLASH\ implementation of the HLLE
scheme \citep{einfeldt88}. The code has been extended with additional
computational modules to handle the radiative losses and the
anisotropic thermal conduction \citep[see][ for the details of the
implementation]{2007A&A...464..753P}.

The 2.5D Cartesian $(x,y)$ grid extends between $-4$ and 4~pc in the $x$
direction and between $-1.4$ and 6.6~pc in the $y$ direction. Initially
the cloud is located at $(x,y) = (0,0)$ and the primary shock front
propagates in the direction of the $y$ axis. At the coarsest resolution,
the adaptive mesh algorithm used in the \FLASH\ code uniformly covers
the 2.5D computational domain with a mesh of $4^2$ blocks, each with
$8^2$ cells. We allow for 5 levels of refinement, with resolution
increasing twice at each refinement level. The refinement criterion
adopted \citep{loehner} follows the changes of the density and of the
temperature. This grid configuration yields an effective resolution of
$\approx 7.6\times 10^{-3}$ pc at the finest level, corresponding to
$\approx 132$ cells per cloud radius. In Sect. \ref{sp_resol}, we
discuss the effect of spatial resolution on our results, considering the
additional runs TR-Bz4-hr and TR-Bz4-hr2 which use an identical setup to
run TR-Bz4, but with higher resolution ($\approx 264$ and $\approx 528$
cells per cloud radius, respectively; see Table \ref{tab2}).

We use a constant inflow boundary condition for the post-shock gas at the
lower boundary, with free outflow elsewhere. For runs with zero magnetic
field ($\beta_{\rm 0} = \infty$), we use reflecting boundary conditions
at $x = 0$ along the symmetry axis of the problem and only evolve half
of the grid.


\section{Results}
\label{sec3}

\subsection{Dynamical evolution}
\label{d_evol}

\begin{figure*}[!h]
  \centering
  \epsscale{1.07}
  \plotone{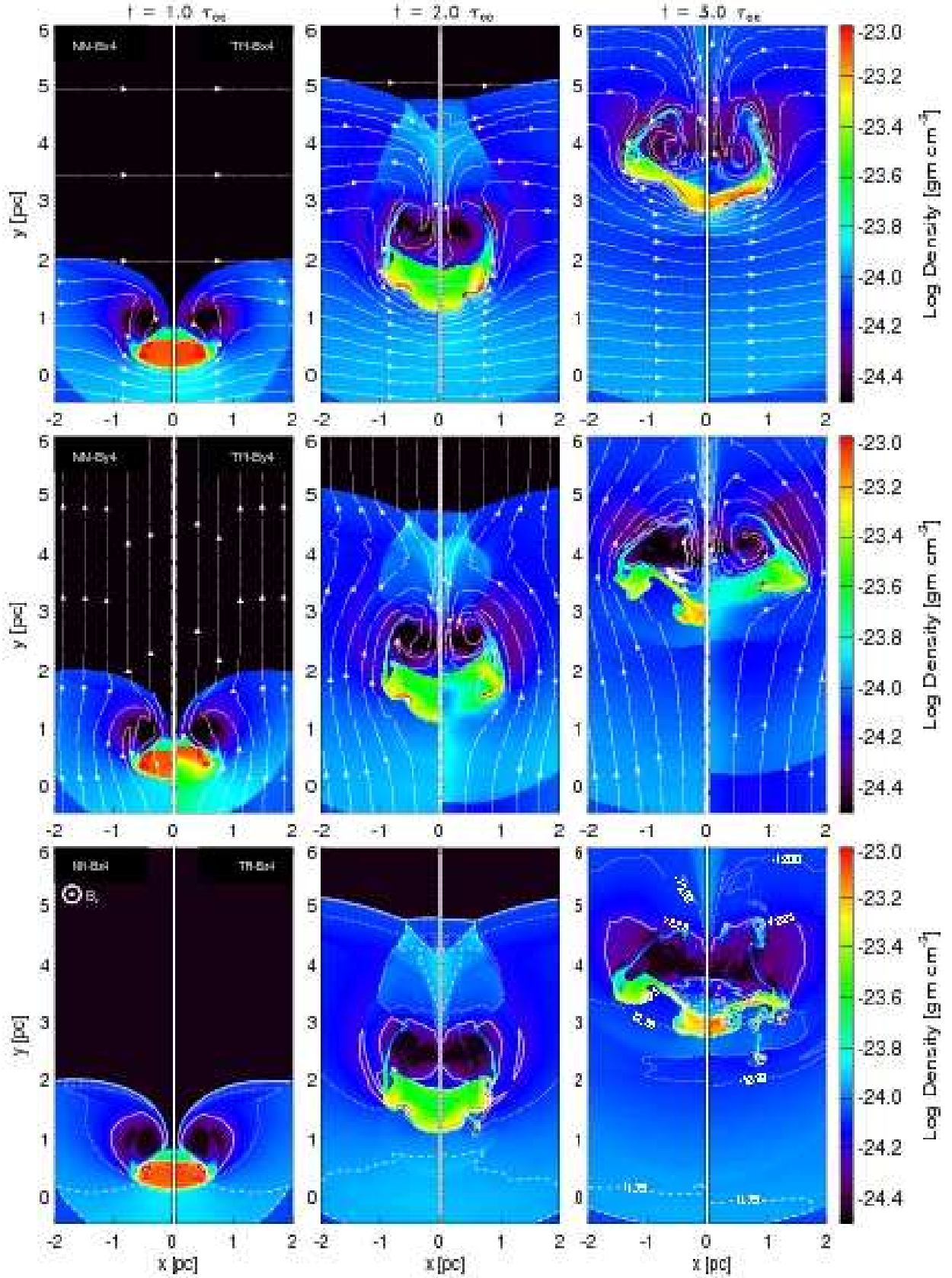}
  \caption{As in Fig.~\ref{fig1} for the simulations with $\beta_0 =
   4$ and the magnetic field oriented along $x$ (upper panels), $y$ (middle
   panels), and $z$ (lower panels). The figure shows the distribution in
   models either without (left half panels) or with (right half panels)
   thermal conduction and radiative losses. For runs NN-Bx4, TR-Bx4,
   NN-By4 and TR-By4, we plot the magnetic field lines; for runs NN-Bz4
   and TR-Bz4, we include contours of $\log (B^2/8\pi)$.}
   \label{fig2}
\end{figure*}

Figs.~\ref{fig1} and \ref{fig2} show the evolution of the mass density in
the $(x,y)$ plane in the simulations with $\beta_0 = \infty$ (runs NN and
TR) and with $\beta_0 = 4$ (runs NN-Bx4, NN-By4, NN-Bz4, TR-Bx4, TR-By4,
TR-Bz4). The left (right) half panels show the result of models without
(with) thermal conduction and radiative losses.

From Fig.~\ref{fig1}, we note that the thermal conduction drives
the cloud evolution in the unmagnetized case ($\beta_0 = \infty$;
run TR): after the initial compression due to the primary shock, the
cloud expands and gradually evaporates due to the heating driven by the
thermal conduction in a few dynamical timescales (see right half panels
in Fig.~\ref{fig1}). The heat conduction strongly contrasts the radiative
cooling of some parts of the cloud and no thermal and hydrodynamic
instabilities (visible in run NN; see left-panels in Fig.~\ref{fig1})
develops during the cloud evolution, making the cloud more stable and
longer-living (the mass mixing is strongly reduced; see Paper I for
more details).

We now discuss the effect of the magnetic-field-oriented (anisotropic)
thermal conduction on the shock-cloud collision when an ambient magnetic
field permeates the ISM. We first summarize the expected evolution in the
presence of an ambient magnetic field, according to the well-established
results of previous models without thermal conduction. We distinguish
between fields perpendicular to the cylindrical clouds (i.e. with
only $B_{\rm x}$ and $B_{\rm y}$ components; referred to as ``external''
fields by \citealt{2005ApJ...619..327F}) and fields parallel to the
cylindrical clouds (i.e. with only the $B_{\rm z}$ component; referred to
as ``internal'' fields). In the former case, the magnetic field plays
a dominant role along the cloud surface and in the wake of the cloud
where it reaches its highest strength (and the plasma $\beta$ its lowest
values; e.g. \citealt{1994ApJ...433..757M}; \citealt{1996ApJ...473..365J}). In
the case of $B_{\rm x}$, the magnetic field is trapped at the nose of
the cloud, leading to a continuous increase of the magnetic pressure
and field tension there (see upper panels in Fig.~\ref{fig2}); in the
case of $B_{\rm y}$, the cloud expansion leads to the increase of magnetic
pressure and field tension laterally to the cloud (see middle panels in
Fig.~\ref{fig2}). In the case of $B_{\rm z}$ (internal field), the magnetic
field, being parallel to the cylindrical cloud, modifies only the total
effective pressure of the plasma (\citealt{1996ApJ...473..365J}); in
the case of radiating shocks, the additional magnetic pressure may play
a crucial role in the shocked cloud, preventing further compression of
the cloud material (\citealt{2005ApJ...619..327F}).

\subsubsection{External magnetic fields}
\label{external_field}

In the case of predominantly external magnetic fields,
\cite{1994ApJ...433..757M} and \cite{1996ApJ...473..365J} have shown
that the hydrodynamic instabilities can be suppressed even in models
neglecting the thermal conduction due to the tension of the magnetic
field lines which maintain a more laminar flow around the cloud surface
(see also \citealt{2005ApJ...619..327F}): for a $\gamma=5/3$ gas,
the KH instabilities are suppressed if $\beta < 2/\mach^2$, whereas
RT instabilities are suppressed if $\beta < (2/\gamma)(\chi/\mach)^2$
(see also \citealt{1961hhs..book.....C}). However, for the parameters
used in this paper ($\chi = 10$ and $\mach = 50$), the magnetic field
cannot suppress KH instabilities in any of our runs, whereas the RT
instabilities are suppressed only in runs that lead to locally very strong
field ($\beta < 0.05$). This can be seen in model NN-Bx4 (upper panels in
Fig.~\ref{fig2}), presenting a large field increase at the cloud boundary,
compared to model NN (Fig.~\ref{fig1}): in the latter case the growth
of KH and RT instabilities at the cloud boundary is much more evident
than in NN-Bx4. On the other hand, the hydrodynamic instabilities are
suppressed more efficiently in models including the thermal conduction
(runs TR-Bx4 and TR-By4) even in cases with low field increase (for
instance in our $B_{\rm y}$ case) as it is evident in Fig.~\ref{fig2}
by comparing models NN-Bx4 and NN-By4 with models TR-Bx4 and TR-By4,
respectively.

\begin{figure*}
  \centering
  \epsscale{1.07}
  \plotone{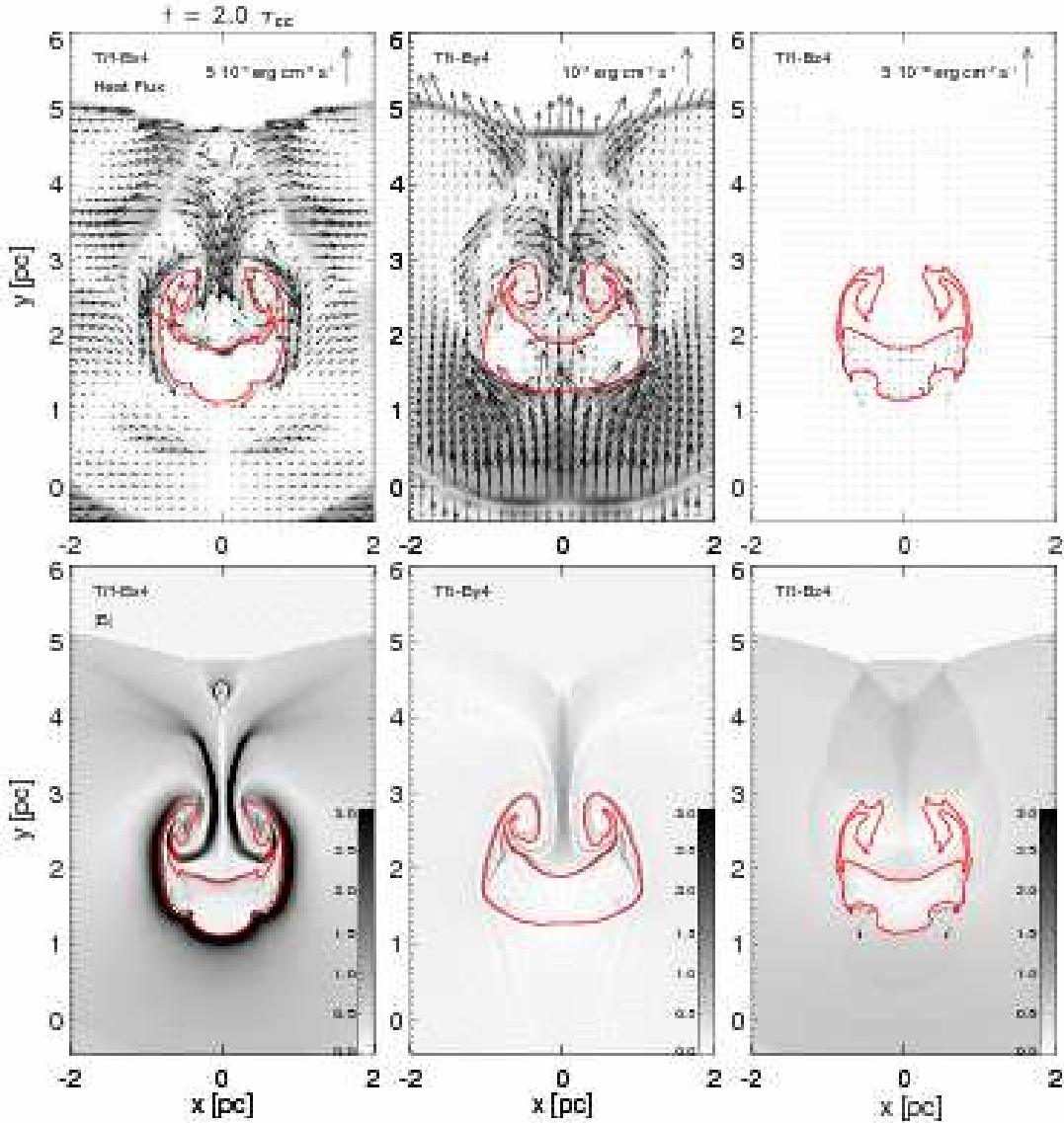}
  \caption{Heat flux (upper panels) and magnetic field
   strength (lower panels) distributions in the $(x,y)$ plane in the
   simulations TR-Bx4 (left panels), TR-By4 (center), and TR-Bz4 (right),
   at time $t=2~\tau_{\rm cc}$. The arrows in the upper panels describe
   the heat flux and scale linearly with respect to the reference value
   shown in the upper right corner of each panel. The scale of the
   magnetic field strength is linear and is given by the bar on the
   right, in units of 10~$\mu$G. The red contour encloses the cloud
   material.}
  \label{fig3}
\end{figure*}

The thermal exchanges between the cloud and the surrounding
medium strongly depend on the initial field orientation.
Fig.~\ref{fig3} shows the heat flux and magnetic field strength
distributions in the $(x,y)$ plane in runs TR-Bx4, TR-By4, and TR-Bz4,
at time $t=2~\tau_{\rm cc}$. In our $B_{\rm x}$ case (upper panels
in Fig.~\ref{fig2}), the magnetic field lines gradually envelope
the cloud, reducing the heat conduction through the cloud surface
(see left panels in Fig.~\ref{fig3}): thermal exchanges
between the cloud and the surrounding medium are channelled through
small regions located at the side of the cloud. The cloud expansion
and evaporation are strongly limited by the confining effect of the
magnetic field (cf. the unmagnetized case TR in Fig.~\ref{fig1} with
model TR-Bx4 in Fig.~\ref{fig2}) that becomes up to 30 times stronger
just outside the cloud than inside it (see, also, the lower left
panel in Fig.~\ref{fig3}). The consequent thermal insulation induces
the radiative cooling and condensation of the plasma into the cloud
during the phase of cloud compression ($t < \tau_{\rm cc}$). At the
end of this phase, the cloud material has temperature $T\approx 10^5$~K
and density $n_{\rm H}\approx 10$~cm$^{-3}$ where primary and reverse
shocks transmitted into the cloud are colliding; for these values of $T$
and $n_{\rm H}$, the Field length scale (\citealt{1990ApJ...358..375B})
derived from the ratio of cooling timescale over conduction timescale
(see Paper I for details) is

\begin{equation}
l \approx 10^6 \frac{T^2}{n_{\rm H}} \approx 3.2\times
10^{-4}\mbox{~pc}~.
\label{ratio}
\end{equation}

\noindent
The radiative cooling dominates over the effects of the thermal conduction
in cold and dense regions with dimensions larger than $l$. At variance
with our unmagnetized case TR, therefore, thermal instabilities develop
in run TR-Bx4. One of this cold and dense structures is evident in
Fig.~\ref{fig2} (upper panels) and is located at the cloud boundary
near the nose of the cloud (at $x\approx 0.4$ pc and $y\approx 3.0$ pc)
at $t=3~\tau_{\rm cc}$.

In the $B_{\rm y}$ case, the initial field direction is mostly maintained
in the cloud core during the evolution, allowing efficient thermal
exchange between the core and the hot medium upwind of the cloud (see
center panels in Fig.~\ref{fig3}): the core is gradually heated and
evaporates in few dynamical timescales. This is illustrated by run
TR-By4 in Fig.~\ref{fig2}. On the other hand, the cloud is thermally
insulated laterally where the magnetic field lines prevent thermal
exchange between the cloud and the surrounding medium. Also, a strong
magnetic field component along the $x$ axis develops in the wake of the
cloud and inhibits thermal conduction with the medium downwind of the
cloud. The thermal insulation at the side of the cloud determines the
growth of thermal instabilities where shocks transmitted into the cloud
collide (see middle panels in Fig.~\ref{fig2}).

In both external field configurations, elongated structures of strong
field concentration are produced on the axis downwind of the cloud
due to the focalization of the magnetized fluid flows there (see
upper and middle panels of Fig.~\ref{fig2}, and lower panels in
Fig.~\ref{fig3}). These filamentary structures, identified as ``flux
ropes'' by \cite{1994ApJ...433..757M}, are formed by magnetic field lines
stretched around the cloud shape and do not carry a significant amount of
cloud material (as shown by the tracer $C_{\rm cl}$) although the plasma
there moves with the cloud (see also \citealt{2000ApJ...543..775G}).

\subsubsection{Internal magnetic fields}

Predominantly internal magnetic fields strongly suppress the heat
conduction, providing an efficient thermal insulation of the cloud
material (see right panels in Fig.~\ref{fig3}). In the realistic
configuration of an elongated cloud with finite length $L$ along the $z$
axis, some heat would be conducted along the magnetic field lines. The
characteristic timescales for the conduction along magnetic field lines is
(see Paper I)

\begin{equation}
\tau_{\rm cond}\approx 2.6\times 10^{-9} \frac{n_{\rm H}L^2}{T^{5/2}}~.
\end{equation}

\noindent
We estimate that the cloud would thermalize in $\tau_{\rm cond}>
3.5~\tau_{\rm cc}$ (i.e. the physical time covered by our simulations),
if the length scale of the cloud along the $z$ axis is $L> 3$~pc. In
this case, hydrodynamic instabilities develop at the cloud boundary, being
both the magnetic field and the thermal conduction not able to suppress
them. The growth of these instabilities is clearly seen in Fig.~\ref{fig2}
(lower panels). The combined effect of hydrodynamic instabilities
and shocks transmitted into the cloud leads to unstable high-density
regions at the cloud boundaries that trigger the development of thermal
instabilities there (see lower panels in Fig.~\ref{fig2}). However,
as discussed by \cite{2005ApJ...619..327F}, internal magnetic field
lines are expected to resist compression in the shocked cloud, thus
reducing the cooling efficiency. In fact, in our run TR-Bz4, the cloud
material is prevented from cooling below $T\approx 10^3$~K. Since the
thermal conduction does not play any significant role in the shock-cloud
interaction, our $B_{\rm z}$ case leads to results similar to those obtained
by \cite{2005ApJ...619..327F} and we do not discuss further this case.

\subsection{Role of thermal conduction}

In this section, we study more quantitatively the effect of thermal
conduction on the cloud evolution and, in particular, on the cloud
compression and on the magnetic field increase. To this end, we use the
tracer defined in Sect. \ref{sec2} to identify zones whose content is
the original cloud material by more than 90\%. Then, we define the
cross-sectional area of cloud material, $A_{\rm cl}(t)$, as the total
area in the $(x,y)$ plane occupied by these zones. We define the cloud
compression (or expansion) as $A_{\rm cl}/A_{\rm cl0}$, where $A_{\rm
cl0}$ is the initial cross-sectional area. We also define an average
mass-weighted temperature of the cloud and an average magnetic field
strength associated to the cloud as

\begin{equation}
\langle T \rangle_{\rm cl} = \frac{\displaystyle
\int_{A(C_{\rm cl}>0.9)} C_{\rm cl}~\rho T ~da}{\displaystyle
\int_{A(C_{\rm cl}>0.9)} C_{\rm cl}~\rho ~da}
\end{equation}

\begin{equation}
\langle B \rangle_{\rm cl} = \frac{\displaystyle
\int_{A(C_{\rm cl}>0.9)} C_{\rm cl}~B ~da}{\displaystyle
\int_{A(C_{\rm cl}>0.9)} C_{\rm cl}~da}
\end{equation}

\begin{figure*}
  \centering
  \epsscale{1.0}
  \plotone{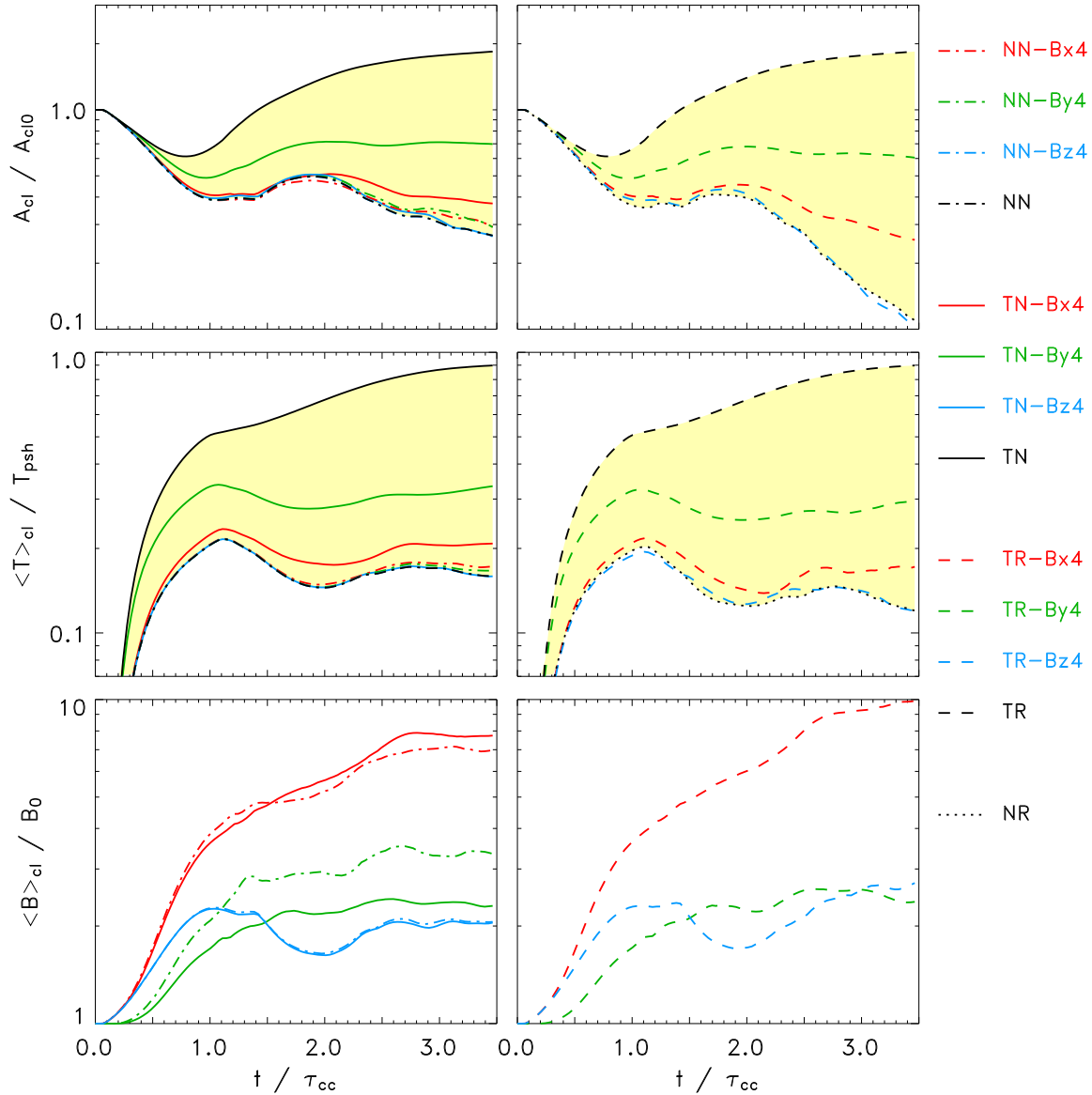}
  \caption{Evolution of cloud compression (upper panels),
   of average temperature (middle panels), and of average magnetic
   field strength (lower panels) of the cloud for runs which neglect the
   thermal conduction and the radiation (dot-dashed lines; left panels),
   for runs which include the thermal conduction but neglect the radiation
   (solid; left panels), for runs which include the radiation but neglect
   the thermal conduction (dotted; right panels) and for runs which
   include both physical effects (dashed; right panels). The magnetized
   cases with $\beta_{\rm 0} = 4$ are marked with red (initial magnetic
   field along $x$), green (initial {\mag} along $y$) and blue (initial
   {\mag} along $z$) lines; the unmagnetized cases are marked with black
   lines. The light yellow regions mark the location of solutions which
   have thermodynamical characteristics in between the cases of maximum
   efficiency of the thermal conduction (models TN and TR) and the
   cases without thermal conduction (models NN and NR). By comparing the
   position of the magnetized models curves inside the yellow region, it
   is possible to quantitatively assess the degree of suppression of the
   effects of the thermal conduction by the magnetic fields.}
  \label{fig4}
\end{figure*}

\noindent
where we integrate on zones with $C_{\rm cl}>0.9$. Note that our
choice of considering cells with the value of the passive tracer $C_{\rm
cl}>0.9$ is arbitrary. To determine how sensitive the results are to this
value and, in particular, to small changes in it, we derive our results
also considering the values $C_{\rm cl}>0.85$ and $C_{\rm cl}>0.95$.
In all the cases, we find that the results derived with the different
thresholds show the same trend with differences lower than $10\%$.

Fig.~\ref{fig4} shows the cloud compression, $A_{\rm cl}/A_{\rm cl0}$,
the average temperature of the cloud, $\langle T \rangle_{\rm cl}$,
normalized to the post-shock temperature of the surrounding medium
($T_{\rm psh}=4.7\times 10^6$ K), and the average magnetic field strength
associated to the cloud, $\langle B \rangle_{\rm cl}$, normalized to
the initial field strength ($B_{\rm 0} = 1.31\;\mu$G, corresponding to
$\beta_{\rm 0}=4$) as a function of time for models neglecting thermal
conduction and radiation (hereafter NNs models), for models including
conduction but neglecting radiation (TNs models), and for models including
both conduction and radiation (TRs models); we also include the results
derived from the unmagnetized case NR with radiative cooling and without
thermal conduction. The figure shows both the magnetized cases with
$\beta_{\rm 0}=4$ and the unmagnetized cases (see Table \ref{tab2}).

In all the NNs models either with (NN-Bx4, NN-By4, and NN-Bz4) or without
(NN) the magnetic field, the evolution of the cloud compression and
of the average cloud temperature is roughly the same (see left panels
in Fig.~\ref{fig4}). The cloud is initially compressed over a timescale
$t\approx \tau_{\rm cc}$ due to the ambient post-shock pressure; during this
phase $\langle T \rangle_{\rm cl}$ rapidly increases. After $t\approx
\tau_{\rm cc}$, the cloud partially reexpands, leading to a decrease of
$\langle T \rangle_{\rm cl}$. In the last phase ($t > 2.0~\tau_{\rm cc}$),
the cloud is compressed again by the interaction with the ``Mach stem''
formed during the reflection of the primary shock at the symmetry axis,
and $\langle T \rangle_{\rm cl}$ increases; later $A_{\rm cl}/A_{\rm cl0}$
continues to decrease, because of the mixing of the cloud material with
the ambient medium (see Sect. \ref{mix}; see also Paper I), while $\langle
T \rangle_{\rm cl}$ stabilizes at $\approx 0.17\;T_{\rm psh}$.

The field increase in the cloud material depends on the initial
configuration of {\mag} (see lower left panel in Fig.~\ref{fig4}). In
the case of external fields ($B_{\rm x}$ and $B_{\rm y}$ components), {\mag}
is mainly intensified due to stretching of field lines due to sheared
motion. In the $B_{\rm x}$ case, the magnetic field undergoes the greatest
increase and $\langle B \rangle_{\rm cl}$ keeps increasing during
the whole evolution. In fact the field is mainly intensified at the nose
of the cloud where the background flow continues to stretch the field
lines during the evolution (see upper panels in Fig.~\ref{fig2}). In the
$B_{\rm y}$ case, the field increase occurs mainly at the side of the
cloud where the field lines are stretched along the cloud surface. In the
case of internal fields ($B_{\rm z}$ component), the field increase
is due to squeezing of field lines through compression. $\langle B
\rangle_{\rm cl}$, therefore, follows the changes in $A_{\rm cl}/A_{\rm cl0}$,
since the field is locked within the cloud material. Thus the greatest
field increase occurs at $t\approx \tau_{\rm cc}$ when the shocks
transmitted into the cloud collide.

The effects of thermal conduction are greatest in the unmagnetized
model (TN) which can be considered an extreme limit case (see left
panels in Fig.~\ref{fig4}). During the first stage of evolution
($t < 0.8\;\tau_{\rm cc}$), the cloud is heated efficiently by the
thermal conduction and its average temperature increases rapidly to
$\sim 0.5~T_{\rm psh}$. As a consequence, the pressure inside the cloud
increases and the cloud reexpands earlier than in model NN. Afterwards,
the average cloud temperature, $\langle T \rangle_{\rm cl}$, keeps
increasing up to $\sim 0.9~T_{\rm psh}$ at $t=3.5\;\tau_{\rm cc}$. 

In the case of predominantly external magnetic fields (models TN-Bx4
and TN-By4), the thermal conduction still plays a significant role in
the cloud evolution, although its effects are not as large as in the
unmagnetized case (TN). During the initial compression, the thermal
conduction contributes to the cloud heating: the average temperature of
the cloud reaches values larger than in models neglecting the conduction
(compare TNs with NNs models in the left panels of Fig.~\ref{fig4}). This
effect is greatest in the $B_{\rm y}$ case which is the configuration
of field lines that allows the most efficient thermal exchange
between the cloud and the hot environment (see Sect. \ref{d_evol}). At
$t=3.5\;\tau_{\rm cc}$, $\langle T \rangle_{\rm cl}$ in TNs models reaches
values larger than in NNs models ($\approx 0.21\;T_{\rm psh}$ in the
$B_{\rm x}$ case and $\approx 0.33\; T_{\rm psh}$ in the $B_{\rm y}$ case).
For internal magnetic fields, the thermal conduction plays no role in
the evolution of the shocked cloud, being strongly ineffective due to
{\mag} (see Sect. \ref{d_evol}). As a consequence, the TN-Bz4 model
leads to the same results as NNs models.

In general, therefore, the effects of the thermal conduction in the
presence of an ambient magnetic field are reduced with respect to the
corresponding unmagnetized case, but not entirely suppressed. This
can be seen in Fig. \ref{fig4}, where we have marked in light yellow
the region between the fully conductive unmagnetized case (TN) and the
case without thermal conduction (NN). The magnetized TNs models are always
within this region, meaning that the effects of the thermal conduction
are never as large as in the unmagnetized case (TN) but not completely
suppressed as in the model NN.

We also note that the thermal conduction influences indirectly the
magnetic field increase. The main changes are in the $B_{\rm y}$ case and
are due to the larger expansion of the cloud that reduces the increase
of the field associated to the cloud, being the field locked within the
cloud material.

In our unmagnetized case TR (including thermal conduction and
radiative cooling), the thermal conduction prevents the onset of
thermal instabilities, and the evolution of the shocked cloud is
the same as found in the TN model. At variance with our unmagnetized
case TR, Fig.~\ref{fig4} shows that thermal instabilities develop in
all our magnetized TRs runs, being the effects of thermal conduction
reduced by the magnetic field. The effects of radiative cooling are
very strong for internal fields (our $B_{\rm z}$ case; see run TR-Bz4
in Fig.~\ref{fig4}). In this case, the heat conduction is totally
suppressed by the magnetic field and the evolution of $A_{\rm cl}/A_{\rm
cl0}$ and of $\langle T \rangle_{\rm cl}$ are the same as those found
in the unmagnetized case with radiative cooling and without thermal
conduction (model NR); at $t=3.5\;\tau_{\rm cc}$, run TR-Bz4 (and NR)
shows the largest cloud compression ($A_{\rm cl}/A_{\rm cl0} \approx
0.1$) and the lowest cloud average temperature ($\langle T \rangle_{\rm
cl}\approx 0.12\;T_{\rm psh}$). In the case of external fields (runs
TR-Bx4 and TR-By4), the effects of heat conduction are reduced but not
suppressed and the results are intermediate between those derived for
runs NR and TR (i.e. within the light yellow region in the right panels
in Fig. \ref{fig4}). The cooling efficiency is largely reduced in our
$B_{\rm y}$ case (run TR-By4), namely that with the magnetic field
configuration that allows the most effective thermal conduction.

\subsection{Mass mixing and energy exchange}
\label{mix}

We use the tracer to derive the cloud mass, $M_{\rm cl}$, as the total
mass in zones whose content is the original cloud material by more
than 90\%,

\begin{equation}
M_{\rm cl} = L~\int_{A(C_{\rm cl}>0.9)} C_{\rm cl}~\rho~da~,
\label{mass}
\end{equation}

\noindent
where $L$ is the cloud length along the $z$ axis, and the integral is
done on zones with $C_{\rm cl}>0.9$. We investigate the mixing of cloud
material with the ambient medium by defining the remaining cloud mass
as $M_{\rm cl}/M_{\rm cl0}$, where $M_{\rm cl0}$ is the initial cloud mass.

The tracer allows us to investigate also the energy exchange between
the cloud and the surrounding medium; we derive the internal energy,
${\cal I}_{\rm cl}$, and the kinetic energy, ${\cal K}_{\rm cl}$ of the
cloud as

\begin{equation}
{\cal I}_{\rm cl} = L~\int_{A(C_{\rm cl}>0.9)} C_{\rm cl}~\rho\epsilon~da~,
\label{eint}
\end{equation}

\begin{equation}
{\cal K}_{\rm cl} = \frac{L}{2} \int_{A(C_{\rm cl}>0.9)} C_{\rm cl}~\rho 
|{\bf u}|^2~da~,
\label{ekin}
\end{equation}

\noindent
where again $L$ is the cloud length along the $z$ axis, and the integral
is done on zones with $C_{\rm cl}>0.9$. We also define the total energy
of the cloud as

\begin{equation}
E_{\rm cl} = {\cal I}_{\rm cl}+{\cal K}_{\rm cl}~.
\end{equation}

\begin{figure*}[!ht]
  \centering
  \epsscale{1.0}
  \plotone{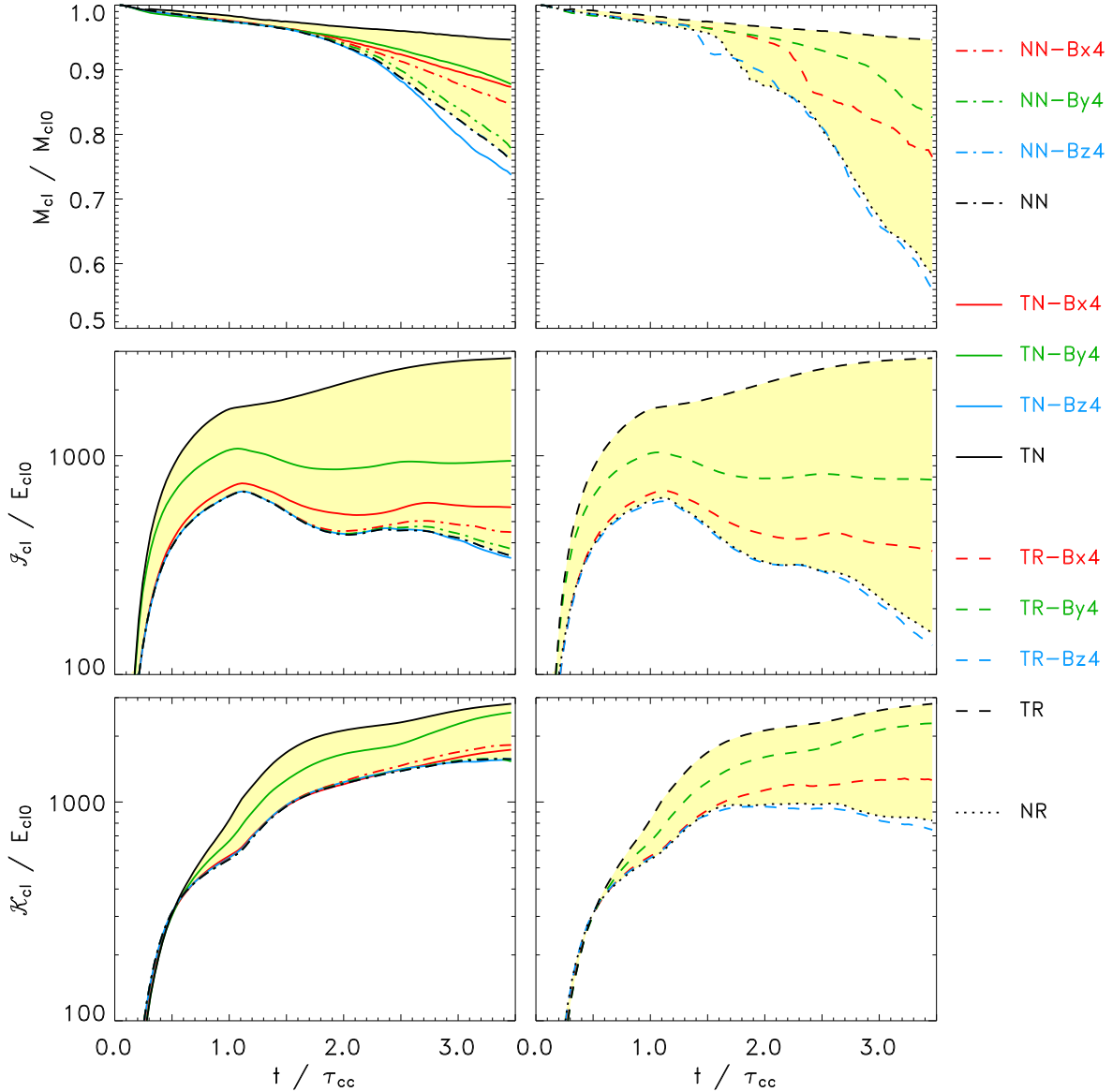}
  \caption{Presentation as in Fig.~\ref{fig4} for the evolution
   of the cloud mass (upper panels), of the internal energy of the
   cloud (middle panels), and of the kinetic energy of the cloud (lower
   panels).}
  \label{fig5}
\end{figure*}

Fig.~\ref{fig5} shows the evolution of the cloud mass, $M_{\rm
cl}/M_{\rm cl0}$, for NNs, TNs, and TRs models; again we also include the
unmagnetized case with radiative cooling and without thermal conduction
(model NR). Both unmagnetized cases and magnetized cases with $\beta_{\rm
0}=4$ are shown. In models without thermal conduction and radiation
(NNs models), the hydrodynamic instabilities drive the mass mixing of
the cloud\footnote{This is also true in our magnetized cases because,
for the parameters used in this paper ($\mach = 50$ and $\chi=10$), the
hydrodynamic instabilities are partially suppressed by the magnetic field
only in runs evolving to strong fields (see Sect. \ref{d_evol}).}. The
mass loss rate of the cloud, $\dot{m}_{\rm cl}$, increases significantly
after $1.5\;\tau_{\rm cc}$ (i.e. after the hydrodynamic instabilities
have fully developed at the cloud boundary), with $\dot{m}_{\rm cl}
\approx 1.5\times 10^{-6} L_{\rm pc}~~M_{\odot}$~yr$^{-1}$, where $L_{\rm
pc}$ is the cloud length along the $z$ axis in units of pc: $\sim 20$\%
of the cloud mass is contained in mixed zones at $t = 3.5\;\tau_{\rm
cc}$. The only exception is run NN-Bx4 ($\sim 15$\% of the cloud mass
is in mixed zones at $t = 3.5\;\tau_{\rm cc}$), being in this case RT
instabilities partially suppressed by the magnetic field (compare run
NN-Bx4 with runs NN-By4 and NN-Bz4 in Fig.~\ref{fig2}).

In TNs models with external magnetic fields (TN-Bx4 and TN-By4), the
mass loss rate of the cloud is less efficient than in NNs models with
$\dot{m}_{\rm cl}\approx 6\times 10^{-7} L_{\rm pc}~~M_{\odot}$~yr$^{-1}$
($\sim 10$\% of the cloud mass is in mixed zones at $t = 3.5\;\tau_{\rm
cc}$). In fact, in these cases the thermal conduction suppresses most
of the hydrodynamic instabilities and the mass loss mainly comes from
the cloud evaporation driven by the thermal conduction rather than from
hydrodynamic ablation. Note that our unmagnetized TN model is an
extreme limit case in which the hydrodynamic instabilities are totally
suppressed by the thermal conduction which drives the cloud mixing; in
this case, the mass loss rate is $\dot{m}_{\rm cl}\approx 1.5\times
10^{-7} L_{\rm pc}~~M_{\odot}$~yr$^{-1}$ ($\sim 5$\% of the cloud mass is
in mixed zones at $t = 3.5\;\tau_{\rm cc}$).

In magnetized TRs models, the onset of thermal instabilities
increases the mass loss rate of the cloud with respect to the
unmagnetized case ($\dot{m}_{\rm cl}$ ranges between $1.5\times
10^{-6} L_{\rm pc}~~M_{\odot}$~yr$^{-1}$ and $4\times 10^{-6} L_{\rm
pc}~~M_{\odot}$~yr$^{-1}$) due to the fragmentation of the cloud in
dense and cold cloudlets. We expect, therefore, that the larger the
amount of cloud mass mixed with the surrounding medium at the end of the
evolution, the more limited the thermal exchange between the cloud and
the hot ambient medium (and, therefore, the greater the efficiency of
radiative cooling). In fact, the upper right panel in Fig.~\ref{fig5}
shows that the mass mixing has the greatest efficiency in run TR-Bz4
(i.e. in the case with the thermal conduction totally suppressed) which
shows a mass loss rate of the cloud similar to that derived from the
unmagnetized NR model. On the other hand, in runs TR-Bx4 and TR-By4,
the mass mixing is intermediate between those derived with runs NR and TR.

Fig.~\ref{fig5} also shows the evolution of internal (middle panels)
and kinetic (lower panels) energy of the cloud, normalized to the
initial total energy of the cloud, $E_{\rm cl0}$. Among the magnetized
cases considered, the greatest values of ${\cal I}_{\rm cl}$ are reached
in our $B_{\rm y}$ case which is the field configuration that allows the
most efficient thermal exchange between the cloud and the environment;
the increase of ${\cal I}_{\rm cl}$ is due to the heat conducted to the
shocked cloud. Also, the $B_{\rm y}$ case leads to the greatest values of
${\cal K}_{\rm cl}$ because the cloud has a larger cross-sectional area
(because of the larger cloud expansion due to the heating driven by heat
conduction; see upper panels in Fig.~\ref{fig4}) and offers, therefore,
a larger surface to the pressure of the shock front responsible of the
cloud acceleration.

\subsection{Role of the initial field strength}

\begin{figure}[!t]
  \centering
  \epsscale{1.1}
  \plotone{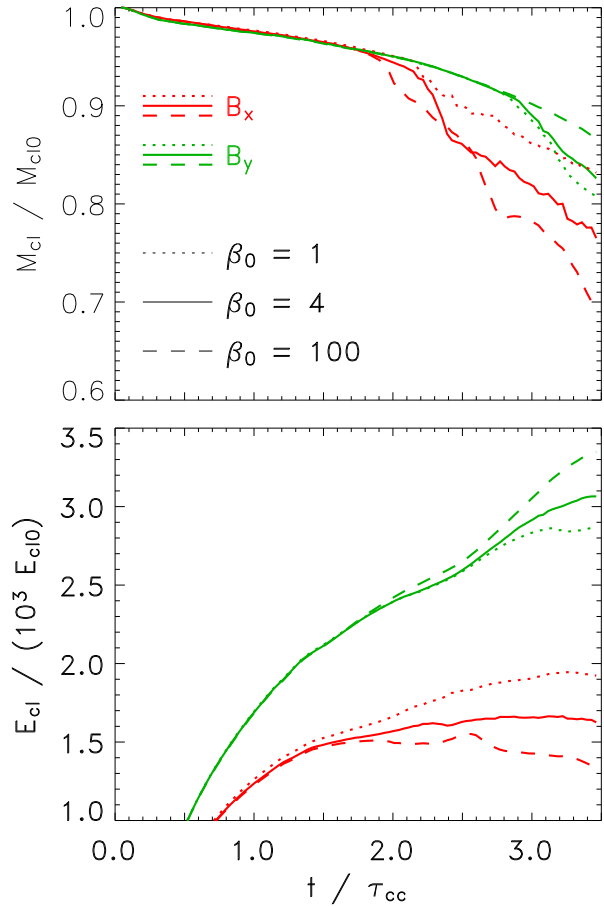}
  \caption{Evolution of the cloud mass (upper panel) and of
   the total energy of the cloud (internal plus kinetic; lower panel) for
   runs including both the thermal conduction and the radiative cooling
   (TRs models). The figure shows the simulations with the magnetic field
   oriented along $x$ (red lines) or $y$ (green) and with $\beta_{\rm 0}
   = 1$ (dotted lines), 4 (solid), and 100 (dashed).}
  \label{fig6}
\end{figure}

In this section we explore the effects of the initial field strength
on the mass mixing and energy exchange of the cloud. Fig. \ref{fig6}
shows the evolution of the cloud mass, $M_{\rm cl}/M_{\rm cl0}$ (upper
panel), and of the total (internal plus kinetic) energy of the cloud,
$E_{\rm cl}/E_{\rm cl0}$ (lower panel), for magnetized TRs models with
different values of $\beta_{\rm 0}$. We discuss here only the cases
of predominantly external magnetic fields ($B_{\rm x}$ or $B_{\rm y}$
case) since no significant dependence on the initial field strength
has been found in the case of predominantly internal magnetic fields
($B_{\rm z}$ case).

Fig. \ref{fig6} shows that the initial field strength plays a significant
role in the $B_{\rm x}$ case. In particular, models with greater values
of $\beta_{\rm 0}$ show a more efficient mixing of the cloud material
and a less rapid increase of the cloud energy. As discussed in Sect.
\ref{mix}, in the $B_{\rm x}$ case, the rate of mass-loss from the cloud is
mainly driven by ablation through the hydrodynamic instabilities (being
the thermal conduction strongly suppressed by the magnetic field). On
the other hand, in the case of external fields, the instabilities can
be dumped by the magnetic field, depending on its strength (see
Sect. \ref{external_field}). For instance, in the $B_{\rm x}$ case with
$\beta_{\rm 0}=4$, we found that the RT instabilities are mostly suppressed
by the magnetic field (see upper panels in Fig. \ref{fig2}). In the
$B_{\rm x}$ case with $\beta_{\rm 0}=100$, instead the magnetic field is
too weak to dump the hydrodynamic instabilities over the timescales
considered; these instabilities, in turn, lead to the formation of
regions dominated by the radiative cooling, triggering the development
of thermal instabilities. Both the hydrodynamic and
the thermal instabilities determine the cloud mass mixing (which is
higher for higher values of $\beta_{\rm 0}$). In addition, the thermal
instabilities reduce the increase of the cloud energy (which is less
rapid for higher $\beta_{\rm 0}$) due to significant radiative losses.

In the $B_{\rm y}$ case, the initial field strength has a smaller influence
on the dynamic and thermal evolution of the cloud than in the $B_{\rm x}$
case (see Fig. \ref{fig6}). In addition, at variance with the $B_{\rm x}$
case, models with greater values of $\beta_{\rm 0}$ show a less efficient
mixing of the cloud material and a more rapid increase of the cloud
energy. In the $B_{\rm y}$ case, in fact, the hydrodynamic instabilities
responsible of the mass mixing are mainly suppressed by the thermal
conduction rather than by the magnetic field as in the $B_{\rm x}$
case. As a consequence, the higher the value of $\beta_{\rm 0}$, the
more effective the thermal conduction in suppressing the instabilities
and in heating the plasma, the less efficient the cloud mass mixing
and the more rapid the increase of the cloud energy.

\subsection{Effect of spatial resolution}
\label{sp_resol}

The effective resolution adopted in our simulations is $\approx
132$ cells per cloud radius, a value above the resolution requirements
suggested by \cite{1994ApJ...420..213K} for non-radiative clouds. However,
for radiative clouds, we expect that the details of the plasma radiative
cooling depend on the numerical resolution: a higher resolution
may lead to different peak density and hence influence the cooling
efficiency of the gas, preventing further compression of the cloud. In
the non-conducting regime, \cite{2005ApJ...619..327F} found that the
results generally converge for simulations with resolution larger than
100 cells per cloud radius ($\lsim 10$\% differences). In the simulations
presented here, the thermal conduction partially contrasts the radiative
cooling in the case of external fields ($B_{\rm x}$ or $B_{\rm y}$),
alleviating the problem of numerical resolution (see also Paper I).

In order to check if our adopted resolution is sufficient to capture the
basic cloud evolution over the time interval considered, we compare three
simulations (TR-Bz4, TR-Bz4-hr, and TR-Bz4-hr2) with different spatial
resolution (132, 264, and 528 zones per cloud radius, respectively) for
the $B_{\rm z}$ case with $\beta = 4$, namely one of the cases in which
the growth of hydrodynamic and thermal instabilities is most prominent
and the effect of thermal conduction (contrasting the development of
hydrodynamic and thermal instabilities) is negligible. Since this case
is one of the most demanding for resolution, it can be considered a
worst case comparison of convergence.

\begin{figure}[!t]
  \centering
  \epsscale{1.1}
  \plotone{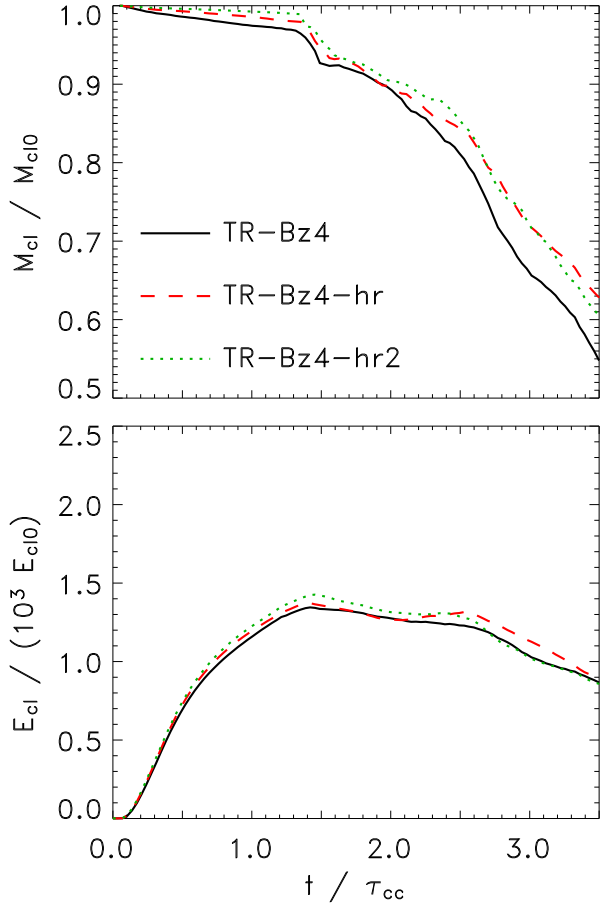}
  \caption{Presentation as in Fig.~\ref{fig6} for runs TR-Bz4
   (solid lines), TR-Bz4-hr (dashed), and TR-Bz4-hr2 (dotted).}
  \label{fig7}
\end{figure}

Figure \ref{fig7} compares the evolution of the cloud mass, $M_{\rm
cl}/M_{\rm cl0}$, and of the total energy of the cloud, $E_{\rm cl}/E_{\rm
cl0}$, for the three simulations TR-Bz4, TR-Bz4-hr, and TR-Bz4-hr2. In
general, we find that the results obtained with the three simulations
agree quite well in their qualitative behavior, showing differences
$\lsim 10\%$. In runs TR-Bz4-hr and TR-Bz4-hr2, the remaining cloud
mass and the total energy of the cloud are, in general, systematically
higher than in run TR-Bz4. The larger mass mixing in TR-Bz4 is driven
by the higher diffusion of the low-resolution grid down to the very
small structures which tend to smear out concentrated density peaks,
promoting mass mixing. The slightly lower energy of the cloud in TR-Bz4 is
a consequence of the larger mass mixing derived in this run with respect
to TR-Bz4-hr and TR-Bz4-hr2. Note that, in runs showing the onset of
thermal instabilities (i.e NRs and TRs models), the size of the latters
reaches the resolution limit toward the end of the simulations when the
relevant physical processes are already at a late stage.

\section{Summary and conclusion}
\label{sec4}

We investigated the importance of magnetic-field-oriented thermal
conduction in the interaction between an isolated elongated dense cloud
and an interstellar shock-wave of an evolved SNR shell through numerical
MHD simulations. To our knowledge, these simulations represent the first
attempt to model the shock-cloud interaction that simultaneously considers
magnetic fields, radiative cooling, and anisotropic thermal conduction.
Our findings lead to several conclusions:

\begin{enumerate}
\item In general, we found that the effects of thermal conduction on the
evolution of the shocked cloud are reduced in the presence of an ambient
magnetic field with respect to the unmagnetized cases investigated in Paper
I. The efficiency of anisotropic thermal conduction strongly
depends on the initial magnetic field orientation and configuration. This
efficiency is the largest when the initial {\mag} is aligned with the
direction of propagation of the shock front, and is the smallest when
{\mag} is aligned with the cylindrical cloud, namely when the heat
conduction is completely suppressed by the magnetic field.

\item We found that the hydrodynamic instabilities are suppressed
efficiently by the anisotropic thermal conduction when the initial
magnetic field is perpendicular to the cylindrical cloud (a configuration
referred to as ``external fields''). On the contrary, in the case
of {\mag} parallel to the cylindrical axis of the cloud (i.e. when
the field has component only along the $z$ axis - internal field),
hydrodynamic instabilities develop at the cloud boundary. We found that,
for the parameters of the simulations chosen, the magnetic tension is
unable to suppress alone the hydrodynamic instabilities.

\item As for thermal instabilities, we found that, depending on the
magnetic field orientation, the heat flux contributes to the heating
of some parts of the cloud, reducing the efficiency of radiative
cooling there, and preventing any thermal instability.

\item The mass loss of the cloud due to mixing with the surrounding
medium is mainly driven by hydrodynamic instabilities; in the case
of external fields (initial {\mag} perpendicular to the cylindrical
cloud) the anisotropic thermal conduction reduces the mass mixing of
the cloud.  In any case, the mass loss rate is larger than that in the
corresponding unmagnetized case ($\dot{m}_{\rm cl}\approx 1.5\times 10^{-7}
L_{\rm pc}~~M_{\odot}$~yr$^{-1}$, i.e. $\sim 5$\% of the cloud mass is in mixed
zones at $t = 3.5\;\tau_{\rm cc}$), but can get very high when the thermal
conduction is completely suppressed ($\dot{m}_{\rm cl}\approx 4\times
10^{-6} L_{\rm pc}~~M_{\odot}$~yr$^{-1}$, i.e. $\sim 45$\% of the cloud mass is
in mixed zones at $t = 3.5\;\tau_{\rm cc}$).

\item The thermal conduction mostly rules the energy exchange between
the cloud and surrounding medium. The exchange is favored when the
magnetic field configuration is such that the conductive flow is not
suppressed (i.e. external field configurations, $B_{\rm x}$ and $B_{\rm y}$
cases), but it is never as high as in the absence of magnetic field. In
the $B_{\rm y}$ case, the cloud core is efficiently heated and evaporates
in few dynamical timescales.

\item In general, the initial magnetic field strength has a small
influence on the dynamic and thermal evolution of the shocked cloud for
the ranges of values explored in this paper (namely $0.26~\mu\mbox{G}
\leq |\mag| \leq 2.63~\mu\mbox{G}$).
\end{enumerate}

It is worth noting that some details of our simulations depend on the
choice of the model parameters. For instance, the onset of thermal
instabilities or the evaporation of the whole cloud depends on the
initial shock Mach number, and on the density and dimensions of the
cloud. The cases that we present here (i.e. $\mach = 50$, $\chi=10$,
and different configurations of \mag) are representative of a regime
in which both the thermal conduction and the radiative cooling play an
important role in the evolution of the shocked cloud. Nevertheless, our
analysis proves that anisotropic thermal conduction can not be neglected
in investigations of the evolution of shocked interstellar clouds.

In our simulations, we consider laminar thermal conduction,
although regions of strong turbulence of different strength and extent
develop in the system (for instance, at the shear layers along the
cloud boundary or at the vortex sheets in the cloud wake). In fact,
the turbulence in these regions may have a significant effect on
thermal conduction, leading to significant deviations of thermal
conductivity from its laminar values (e.g. \citealt{2001ApJ...562L.129N};
\citealt{2006ApJ...645L..25L}); in some cases, the turbulence may
enhance the heat transfer, exceeding the classical Spitzer value
(\citealt{2006ApJ...645L..25L}). As a result, thermal conduction
may be not only anisotropic (in the presence of the magnetic field)
but also ``inhomogeneous'' due to the presence of turbulence. However,
even modeling accurately the turbulent thermal conductivity, we do not
expect significant changes in the results of our $B_{\rm z}$ case, being
the thermal conduction strongly ineffective in the whole spatial domain;
in the remaining cases ($B_{\rm x}$ and $B_{\rm y}$), our modeled thermal
conductivity could be underestimated in regions of strong turbulence,
affecting some details of the simulations but not the main conclusion
of the paper that, in general, anisotropic thermal conduction can play
an important role in the evolution of the shocked cloud.

Note also that the field configurations studied in this work
are highly idealized. More realistic fields are expected to have
more complex topologies and, often, the field can be tangled and
chaotic. In the latter case, the thermal conduction will approach
isotropy, whereas the effect of MHD turbulence is expected to
partially suppress the heat transfer within a factor $\sim 5$
below the classical Spitzer estimate\footnote{As already discussed,
the MHD turbulence can even enhance the heat transfer in some cases
(see \citealt{2006ApJ...645L..25L}).} (\citealt{2001ApJ...562L.129N};
\citealt{2006ApJ...645L..25L}). The shock-cloud collision in the presence
of an organized ambient magnetic field, discussed here, and that in
the absence of magnetic field can be considered as extreme cases: the
former leading to highly anisotropic thermal conduction, the latter to
the classical Spitzer thermal conduction. The case of chaotic magnetic
field is expected to fall in between these two.

Our simulations were carried out in 2.5D Cartesian geometry, implying
that the modeled clouds are elongated along the $z$ axis. This choice
is expected to affect some details of the simulations but not our main
conclusions. Adopting a 3D Cartesian geometry and modeling a spherical
cloud, the highly symmetric shock transmitted into the cloud converging
on the symmetry axis would lead to compression stronger than those found
in our 2.5D simulations, enhancing the radiative cooling. Also, 3D
simulations would provide an additional degree of freedom for hydrodynamic
instabilities, increasing the mass loss rate of the cloud in the cases in
which the mass mixing of cloud material is driven by instabilities. Note
that, for a spherical cloud, our $B_{\rm x}$ and $B_{\rm z}$ cases no
longer differ.

Finally we assume, in our simulations, that the cloud and the
ambient material have the same composition, implying that microscopic
mass mixing due to shear instabilities would be irrelevant. In a more
realistic condition, a cold dense cloud may have a different composition
from the hot ambient flow and the degree of microscopic mixing may
translate into different spectral signatures of the system. In this case,
species diffusion could also be important, along with thermal conduction,
to determine the degree of microscopic mixing of the materials and,
consequently, one would have to ask about the typical values of the
Lewis number (i.e. the ratio of thermal diffusivity to mass diffusivity)
in the system.

It is worth emphasizing that the quantitative results of our
simulations depend on the physical parameters of the model (shock Mach
number, density contrast and dimension of the cloud, etc.) as well as on
the basic assumptions of the model (geometry of the cloud, geometry of
the ambient magnetic field, laminar thermal conduction, composition
of the cloud and of the ambient medium, etc.). Nevertheless, our results
undoubtedly show that the magnetic-field-oriented thermal conduction can
play an important role in the evolution of the shock-cloud interaction
(which depends on the magnetic field orientation and configuration) and,
in particular, in the mass and energy exchange between the cloud and the
hot surrounding medium. We conclude, therefore, that a self-consistent
and quantitative description of the interaction between magnetized
shock-waves and interstellar gas clouds should include the effects of
thermal conduction.

The results presented here are interesting for the study of
middle-aged SNR shells expanding into a magnetized ISM and
whose morphology is affected by ISM inhomogeneities (for
instance, G272.2-3.2, e.g. \citealt{1996rftu.proc..247E};
Cygnus Loop, e.g. \citealt{2002AJ....124.2118P}; Vela SNR, e.g.
\citealt{2005A&A...442..513M}). It will be further interesting to extend
the present study, by modeling the shock-cloud interaction in 3D with
radiative cooling, anisotropic thermal conduction, and magnetic field
included and, even, considering detailed comparisons of model results
with observations.

\acknowledgements
The authors thank Timur Linde for his help with the MHD portion of FLASH
and the referee for constructive and helpful criticism. The software
used in this work was in part developed by the DOE-supported ASC /
Alliance Center for Astrophysical Thermonuclear Flashes at the University
of Chicago, using modules for thermal conduction and optically thin
radiation built at the Osservatorio Astronomico di Palermo. Most of the
simulations have been executed at CINECA (Bologna, Italy) in the framework
of the INAF-CINECA agreement on ``High Performance Computing resources for
Astronomy and Astrophysics''.  This work makes use of results produced by
the PI2S2 Project managed by the Consorzio COMETA, a project co-funded
by the Italian Ministry of University and Research (MIUR) within the
Piano Operativo Nazionale ``Ricerca Scientifica, Sviluppo Tecnologico,
Alta Formazione'' (PON 2000-2006); more information is available at
http://www.pi2s2.it and http://www.consorzio-cometa.it. This work was
supported in part by Istituto Nazionale di Astrofisica.

\bibliographystyle{apj}
\bibliography{references}

\clearpage

\end{document}